\documentclass[10pt,twocolumn,english,twocolumn,aps]{revtex4-1}
\usepackage[latin9]{inputenc}
\usepackage{textcomp}
\usepackage{amssymb}
\usepackage{graphicx}

\makeatletter
\usepackage{lmodern}
\usepackage[T1]{fontenc}
\usepackage{hyperref}

\makeatother

\usepackage{babel}
\begin{document}

\title{Octave-spanning dissipative Kerr soliton frequency combs in $\mathrm{Si_{3}N_{4}}$
microresonators }

\author{Martin H. P. Pfeiffer\textsuperscript{1}, Clemens Herkommer\textsuperscript{1,2},
Junqiu Liu\textsuperscript{1}, Hairun Guo\textsuperscript{1}, Maxim
Karpov\textsuperscript{1}, Erwan Lucas\textsuperscript{1}, Michael
Zervas\textsuperscript{1}, Tobias J. Kippenberg\textsuperscript{1}}
\email{tobias.kippenberg@epfl.ch}

\selectlanguage{english}%

\affiliation{1. \'Ecole Polytechnique F\'ed\'erale de Lausanne (EPFL), CH-1015 Lausanne,
Switzerland}

\affiliation{2. Technische Universit\"at M\"unchen (TUM), D-80333, M\"unchen, Germany}

\begin{abstract}
Octave-spanning, self-referenced frequency combs are applied in diverse
fields ranging from precision metrology to astrophysical spectrometer
calibration. The octave-spanning optical bandwidth is typically generated
through nonlinear spectral broadening of femtosecond pulsed lasers.
In the past decade, Kerr frequency comb generators have emerged as
novel scheme offering chip-scale integration, high repetition
rate and bandwidths that are only limited by group velocity dispersion.
The recent observation of Kerr frequency combs operating in the dissipative
Kerr soliton (DKS) regime, along with dispersive wave formation, has
provided the means for fully coherent, broadband Kerr frequency comb
generation with engineered spectral envelope. Here, by carefully
optimizing the photonic Damascene fabrication process, and dispersion
engineering of $\mathrm{Si_{3}N_{4}}$ microresonators with $1\,\mathrm{THz}$
free spectral range, we achieve bandwidths exceeding one octave at
low powers ($\mathcal{O}(100\,\mathrm{mW})$)  for pump lasers residing
in the telecom C-band ($1.55\,\mathrm{\mu m}$), as well as in the O-band ($1.3\,\mathrm{\mu m}$). Precise dispersion
engineering enables emission of two dispersive waves, increasing the
power in the spectral ends of the comb, down to a wavelength as short
as $850\,\mathrm{nm}$. Equally important, we find that for THz repetition
rate comb states, conventional criteria applied to identify DKS comb
states fail. Investigating the coherence of generated, octave-spanning
Kerr comb states we unambiguously identify DKS states using a response
measurement. This allows to demonstrate octave-spanning DKS comb states
at both pump laser wavelengths of $1.3\mathrm{\,\mu m}$ and $1.55\,\mathrm{\mu m}$
including the broadest DKS state generated to date, spanning more
than $200\,\mathrm{THz}$ of optical bandwidth. Octave-spanning DKS
frequency combs can form essential building blocks for metrology or
spectroscopy, and their operation at $1.3\mathrm{\,\mu m}$ enables
applications in biological and medical imaging such as Kerr comb based optical coherence
tomography or dual comb coherent anti-stokes Raman scattering.
\end{abstract}
\maketitle

\section{Introduction}

\begin{figure}[th]
\includegraphics{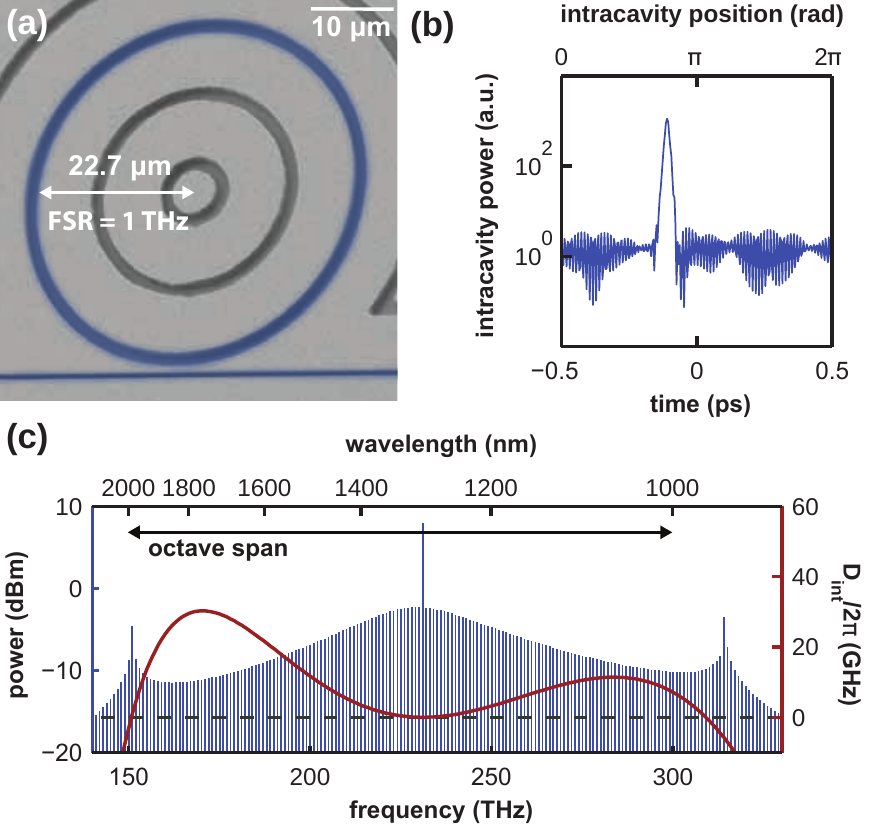}

\caption{Octave-spanning dissipative Kerr soliton frequency comb generation
in a dispersion engineered $\mathrm{Si_{3}N_{4}}$ microresonator.
(a) Scanning electron microscope image of a $\mathrm{Si_{3}N_{4}}$
microresonator with $1\,\mathrm{THz}$ free spectral range during
fabrication. The waveguide structures are highlighted in blue. (b)
Simulated temporal intracavity power profile of an octave-spanning
DKS frequency comb excited with $75\,\mathrm{mW}$ pump power for
$\kappa/2\pi=200\,\mathrm{MHz}$ using the Lugiato-Lefever equation.
The two dispersive waves form an overlapping trailing and leading
wave pattern. (c) The according spectral intracavity power distribution
(blue) and the integrated dispersion profile $\mathrm{D_{int}/2\pi}$
(red) engineered for dispersive wave emission around $1\,\mathrm{\mu m}$
and $2\,\mathrm{\mu m}$ wavelengths. \label{figSimulation}}

\end{figure}

Optical frequency combs with coherent optical bandwidths of one octave
or more are required for many applications, such as precision spectroscopy
\citep{Udem2002}, optical frequency synthesis \citep{Holzwart2000}
or astrophysical spectrometer calibration \citep{Steinmetz2008}.
Conventionally these spectra are synthesized by nonlinear spectral
broadening of a pulsed laser \citep{Udem2002}. However, ensuring
coherent spectral broadening, a smooth spectral envelope, and sufficient
power in the spectral ends for a given pulse source can be challenging, in particular for high repetition rate pulse sources.
Kerr frequency combs \citep{DelHaye2007} have emerged as alternative
scheme that enables compact form factor, high repetition rates and
broadband optical frequency combs, that are even amenable to wafer-scale
integration with additional electrical or optical functionality. The
spectral bandwidth of Kerr frequency combs, being independent of a
specific material gain and primarily determined by the resonator's
group velocity dispersion (GVD), has reached octave span shortly after
the principle's first demonstration \citep{DelHaye2011,Okawachi2011},
although in the high noise operation regime \citep{Herr2012}. Only
recently, the observation of dissipative Kerr soliton (DKS) formation
in microresonators has enabled the controlled excitation of fully
coherent Kerr frequency combs \citep{Herr2013a}. 

When operated in the single soliton state, DKS frequency combs feature
high coherence across their bandwidth and a smooth spectral envelope
that can be numerically predicted with high accuracy using the Lugiato-Lefever
equation or in frequency domain via coupled mode approaches \citep{Coen2013,Okawachi2014,Herr2013a}.
Furthermore, higher order GVD causes soliton Cherenkov radiation,
an oscillatory tail in the temporal soliton pulse profile, corresponding
to a dispersive wave (DW) in the spectral domain, which extends the
spectral comb bandwidth into the normal GVD regime \citep{Akhmediev1995}.
Soliton Cherenkov radiation in microresonators is a fully coherent
process and has e.g. allowed self-referencing and stabilization without
additional spectral broadening via the $2f$-$3f$-method \citep{Brasch2017}.
These properties have made DKS states the preferred operational low
noise states \citep{Herr2012,Xue2015,Saha:13} of Kerr frequency
comb generation with a growing number of applications having been
demonstrated, including e.g. terabit coherent communications \citep{Marin-Palomo2016},
dual comb spectroscopy \citep{Dutt2016,Suh2016} and low noise microwave
generation \citep{Liang2015a}. So far DKS formation has been observed
in a variety of resonator platforms \citep{Brasch2016,Herr2013a,Joshi2016,Liang2015a,Yi2015,Wang2016}
among which planar silicon nitride ($\mathrm{Si_{3}N_{4}}$) waveguide
resonators have gained significant attention. Allowing CMOS-compatible,
wafer-scale fabrication and exhibiting low linear and nonlinear optical
losses in the telecom wavelength region, $\mathrm{Si_{3}N_{4}}$ microresonators
bear realistic potential for applications \citep{Moss2013}. The accurate
control of waveguide dimensions during microfabrication is prerequisite
to precise GVD engineering and thus tailoring of the Kerr frequency
comb bandwidth \citep{Okawachi2014}. Dual dispersive wave emission
extending the spectral bandwidth to both sides is attractive for low
power octave-spanning DKS generation but particularly challenging
to realize as it requires control of higher order GVD. Most microresonators
used for DKS generation to date, did not have specifically engineered
GVD, resulting in limited spectral bandwidth, far below one octave. 

Here we present $\mathrm{Si_{3}N_{4}}$ microresonators with a free
spectral range (FSR) of $1\,\mathrm{THz}$ which allow for DKS frequency
comb generation with bandwidths exceeding one octave. The wide FSR
reduces the total power requirements to cover one octave bandwidth and
benefits the nonlinear conversion efficiency in the DKS state \citep{Bao2014}.
Such microresonators are thus also of interest for wavelength regions
where no high power pump sources exist. We use the recently developed
photonic Damascene process \citep{Pfeiffer2016} for wafer-scale fabrication
of such high-Q microresonator devices with high yield. Figure \ref{figSimulation}(a)
shows a scanning electron microscope (SEM) picture of the microresonator
device including a straight bus waveguide. For small microresonator
radii ($r\thickapprox23\,\mathrm{\mu m}$) the coupling section forms
a relatively large fraction of the total circumference and can cause
parasitic loss. Thus, void-free fabrication of narrow coupling gaps,
as provided by the photonic Damascene process, is important. Moreover,
the bus waveguide cross section needs to be engineered for high ideality
coupling \citep{Pfeiffer2016a}. The fundamental mode families of
the microresonator devices used in this work have a typical internal
linewidth of $\kappa_{0}/2\pi=100\,\mathrm{MHz}$ and thus a loaded
finesse of $\mathcal{F}\thickapprox10^{4}$ at critical coupling.

While a high microresonator finesse is beneficial to lower the power
threshold of Kerr comb generation, the microresonator GVD determines
mostly the power requirements to reach a certain spectral bandwidth.
Microresonator GVD is conveniently expressed in its integrated form
around a central pump frequency $\omega_{0}$:

\[
D_{\mathrm{int}}(\mu)\equiv\omega_{\mu}-(\omega_{0}-D_{1}\mu)=\frac{D_{2}\mu^{2}}{2!}+\frac{D_{3}\mu^{3}}{3!}+\ldots
\]
$D_{\mathrm{int}}(\mu)$ is the resonance frequency deviation of the
$\mathrm{\mu}$-th mode $\omega_{\mu}$ relative to the central pump
mode $\omega_{0}$ from the equidistant grid defined by the repetition
rate $D_{1}/2\pi$. DKS formation requires anomalous GVD ($D_{2}>0$)
as the associated resonance frequency deviation is compensated by
the nonlinear phase shift induced by the soliton. In the case of dominant
quadratic GVD, the $3\,\mathrm{dB}$ bandwidth of the resulting characteristic
$\mathrm{sech^{2}}$ soliton spectral envelope, and thus the temporal
pulse width, is a function of the cavity detuning
and $D_{2}$ \citep{Herr2013a}. Low positive values of $D_{2}$ result
in larger optical bandwidths for a given pump power and detuning.
Beyond this, dispersive wave formation offers an effective way to
coherently enlarge the bandwidth into the normal GVD with locally
enhanced comb tooth power in the spectral ends \citep{Brasch2016,Milian2014}.
Dispersive waves are emitted in resonances which are phase matched
by the soliton induced nonlinear phase shift and the spectral positions
of DWs are thus approximately given by the linear phase matching criterion
$D_{\mathrm{int}}=0$. For dominant, negative fourth-order dispersion
$D_{4}$ two DW positions can be engineered, causing dual dispersive
wave emission.

\begin{figure*}
\includegraphics{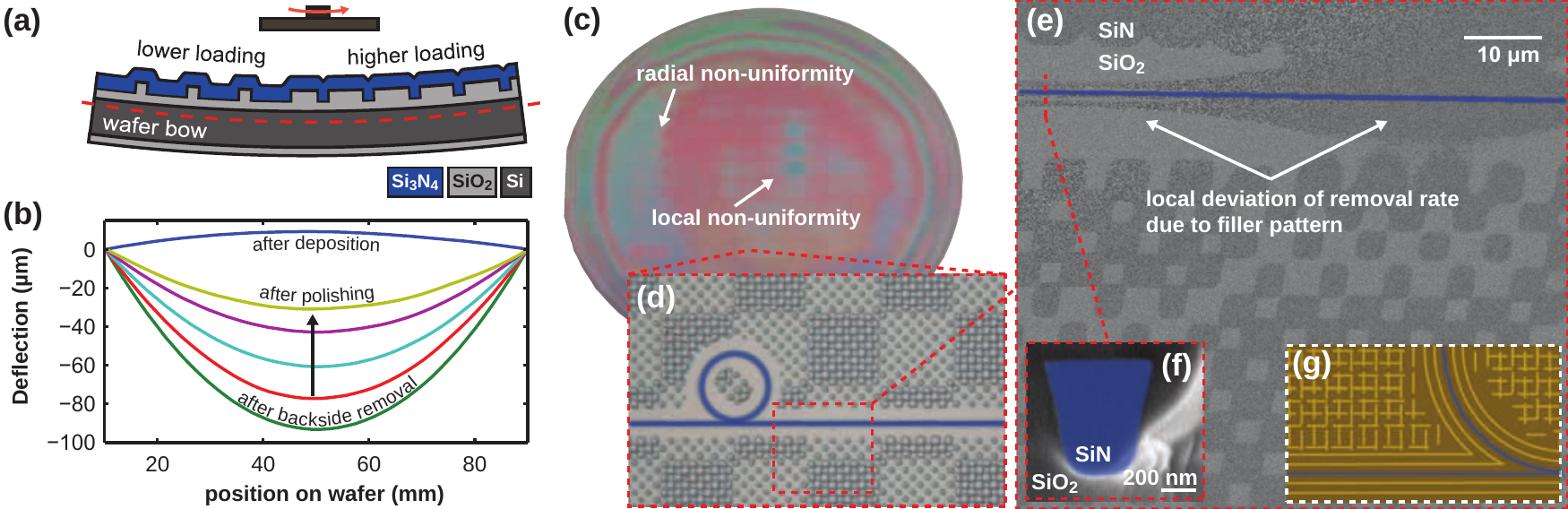}

\caption{Origins of non-uniform planarization in the photonic Damascene process.
(a) Schematic representation of wafer bow and local loading causing
non-uniformity during chemical mechanical planarization. (b) Measurement
of wafer bow evolution during planarization for a $525\,\mathrm{\mu m}$
thick 4'' wafer with $\sim1\,\mathrm{\mu m}$ thick $\mathrm{Si_{3}N_{4}}$
thin film. (c) Photograph of a 4'' wafer after planarization showing
non-uniformity through colored interference patterns. (d) Optical
microscope image of a $1\,\mathrm{THz}$ FSR microresonator surrounded
by a ``checkerboard'' filler pattern. The $\mathrm{Si_{3}N_{4}}$
waveguides are highlighted in blue. (e) SEM image of a non-uniformly
planarized bus waveguide (indicated in blue). The neighboring filler
pattern patches have different loading and thus cause more $\mathrm{Si_{3}N_{4}}$
(dark shaded areas) to remain over the bus waveguide in certain areas.
(f) Cross section of the bus waveguide used to couple light in a $1\,\mathrm{THz}$
FSR microresonator. The dimensions ($\mathrm{0.5\,\mathrm{\mu m}\times0.67\,\mu m}$)
are chosen to provide high ideality coupling to the microresonator's
fundamental $\mathrm{TE_{00}}$ mode family. (g) Optical microscope
image of waveguide structures surrounded by optimized filler pattern.\label{figFabChallenge}}
\end{figure*}

We design the microresonator waveguide cross section based on finite
element (FEM) simulations of the resonator's $\mathrm{TE_{00}}$ mode
family dispersion. Dispersive wave formation is engineered to occur
around wavelengths of $1\mathrm{\,\mu m}$ and $2\mathrm{\,\mu m}$
spanning an octave bandwidth around the pump line at $1.3\,\mathrm{\mu m}$
or $1.55\,\mathrm{\mu m}$. Full 3D finite difference time domain
simulations are employed to engineer the bus waveguide cross section
for high ideality coupling to the $\mathrm{TE_{00}}$ mode family
\citep{Pfeiffer2016a}. Next, we simulate the octave-spanning DKS
generation using a Lugiato-Lefever model \citep{Coen2013,Lugiato1987}
for a critically coupled $\mathrm{Si_{3}N_{4}}$ microresonator with
linewidth $\kappa/2\pi=200\,\mathrm{MHz}$ driven by a pump laser
of $75\,\mathrm{mW}$ power. Figure \ref{figSimulation}(b), (c) show
the intracavity field and the spectral envelope of the simulated single
DKS state. As can be seen the two dispersive waves are present in
the temporal intracavity field as trailing and following wave patterns
which fill the complete cavity. Furthermore, close examination reveals
that the position of the higher frequency dispersive wave is slightly
offset from the linear phase-matching point, indicating the approximative
nature of the criterion $D_{\mathrm{int}}(\mu)=0$.

In the following, we first discuss the challenges of precise waveguide
dimension control using the photonic Damascene process. Leveraging
optimized fabrication procedures, we achieve control over the dispersive
wave positions resulting in octave-spanning comb generation based
on $1.3\mathrm{\,\mu m}$ and $1.55\mathrm{\,\mu m}$ pump lasers.
The coherence of the generated Kerr frequency combs is studied and
a response measurement is used to unambiguously identify DKS formation.
Finally, we study octave-spanning single DKS formation for $1.3\,\mathrm{\mu m}$
and $1.55\,\mathrm{\mu m}$ pump wavelengths featuring the \emph{broadest}
Kerr soliton frequency comb generated to date, exceeding $\mathrm{200\,THz}$
in bandwidth.

\section{Precision dispersion engineering using the photonic Damascene process}

\begin{figure}
\includegraphics{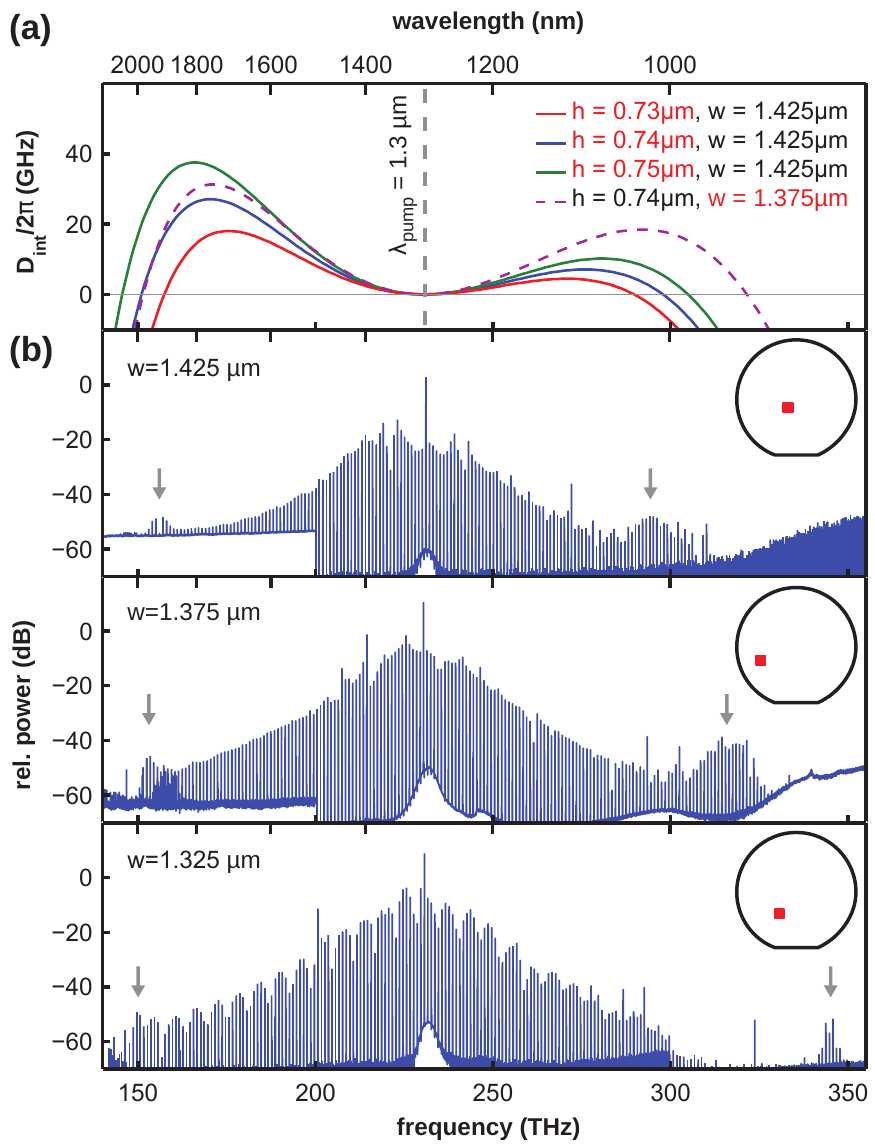}

\caption{Wafer-scale dispersion engineering of octave-spanning Kerr frequency
combs. (a) Simulated integrated dispersion $\mathrm{D_{int}(\nu)/2\pi}$
of fully cladded $\mathrm{Si_{3}N_{4}}$ microresonators with 82\textdegree{}
sidewall angle and different widths and heights for generating dual
dispersive wave DKS states. For a pump wavelength of $1.3\mathrm{\,\mu m}$
dispersive waves can be excited in a linear approximation at positions
close to the zero-points of the integrated dispersion. (b) Unstable
modulation instability, high-noise Kerr frequency combs generated
in three different samples fabricated on the same wafer at different
positions which are indicated in the inset. The spectral position
of the two dispersive waves are indicated by gray arrows. The samples
have by design a difference of $50\,\mathrm{nm}$ in waveguide width
which enables tuning of the high frequency dispersive wave position,
demonstrating the excellent dimension control of the Damascene process.\label{figWaferDispersion}}
\end{figure}

Dispersion engineering for octave-spanning Kerr frequency comb generation
requires stringent waveguide width and height control on the order
of $10\,\mathrm{nm}$, that is challenging to meet. As detailed below
via simulations (shown in Figure \ref{figWaferDispersion}), especially
the position of the dispersive wave features is very sensitive to
variations of the waveguide dimensions. Although DKS frequency comb
generators have been fabricated using conventional subtractive processing,
several challenges arise for the fabrication of high confinement $\mathrm{Si_{3}N_{4}}$
waveguides for nonlinear applications. In particular, these are cracking
of the highly stressed $\mathrm{Si_{3}N_{4}}$ thin film and void
formation at the microresonator coupling gap \citep{Pfeiffer2016}.
The recently reported photonic Damascene process, employed in the
present work, solves these problems by depositing the $\mathrm{Si_{3}N_{4}}$
thin film on a prestructured substrate and subsequent removal of the
excess material using chemical mechanical planarization (CMP). A
dense filler pattern surrounding the waveguide structures effectively
lowers the thin film's tensile stress and enables crack free fabrication
even for micron thick waveguides. Moreover, the filler pattern helps
to control and unify the material removal rate across the wafer during
CMP.

Although the level of control on the waveguide width is similar for
Damascene and subtractive processing schemes (both relying on the
precision of lithography and etching processes) the waveguide height
control is substantially eased in the subtractive fabrication scheme:
a typical thin film deposition non-uniformity of 3\% across a 4''
wafer results in a maximal $25\,\mathrm{nm}$ height deviation for
$800\,\mathrm{nm}$ thick $\mathrm{Si_{3}N_{4}}$ waveguides. For
the Damascene process the variation of the final waveguide thickness
depends on the uniformity of the dry etch process used to structure
the waveguide trenches and most importantly on the uniformity of the
CMP process. While optical interferometry can give precise local information
on the removal rate during CMP, especially the wafer bow and the local
loading lead to height variations (see Figure \ref{figFabChallenge}(a)).
This reduces the yield of devices with dimensions within the tolerance
range for octave-spanning Kerr comb generation, and causes uncontrolled
spectral positions of the dispersive waves.

Figure \ref{figFabChallenge}(b) shows the evolution of the bow during
planarization of a $525\,\mathrm{\mu m}$ thick 4'' wafer measured
using a thin film stress measurement tool. After deposition, the continuous
$\mathrm{Si_{3}N_{4}}$ film on the wafer backside causes a positive
deflection as its tensile stress  is higher than the stress on the
wafer front side which is reduced by the prestructured surface. The
removal of the backside $\mathrm{Si_{3}N_{4}}$ prior to the CMP inverts
the bow to more than $-90\,\mathrm{\mu m}$. The subsequent planarization
step removes the excess $\mathrm{Si_{3}N_{4}}$ on the front side
and continuously relaxes the wafer bow. This bow variation leads to
non-uniformity with radial symmetry visible as the colored interference
ring pattern in Figure \ref{figFabChallenge}(c). Furthermore, local
non-uniformity in the central wafer region can also be observed, which
originates from different removal rates due to loading variations.
Loading refers here to the amount of excess $\mathrm{Si_{3}N_{4}}$
that is in direct contact with the polishing pad and depends on the
local structure density. Such local variations can also occur on much
smaller scales. This is exemplified in Figure \ref{figFabChallenge}(e)
for an area around a bus waveguide (indicated in blue) after CMP.
Depending on the loading of the neighboring filler pattern, more or
less $\mathrm{Si_{3}N_{4}}$ (darker shaded area) is remaining above
the silicon dioxide ($\mathrm{SiO_{2}}$) preform (lighter shaded
area). Such loading dependent non-uniformity leads to local variations
of the waveguide height, and therefore needs to be minimized. Finally,
during landing when most excess $\mathrm{Si_{3}N_{4}}$ is removed
and mostly the $\mathrm{SiO_{2}}$ preform is in contact with the
polishing pad, the material removal rate can drastically change. Thus,
the polishing endpoint is hard to predict and the overall mean waveguide
height differs from the target value.

Here we apply an optimized CMP process that uses higher pressure to
equalize the wafer bow and a thicker wafer substrate ($700\,\mathrm{\mu m}$)
that reduces the amount of initial wafer bow to $-50\,\mathrm{\mu m}$.
The removal rates for $\mathrm{Si_{3}N_{4}}$ and $\mathrm{SiO_{2}}$
are chosen to be similar, limiting the uncertainty during landing.
In order to reduce local loading effects, a novel filler pattern (see
Figure \ref{figFabChallenge}(g)) is used which homogenizes the loading
while providing sufficient stress release for crack-free fabrication.
The dispersion control achieved using this optimized CMP process is
summarized in Figure \ref{figWaferDispersion}. Based on FEM simulations
of microresonator GVD shown in Figure \ref{figWaferDispersion}(a),
the target waveguide height of $0.74\,\mathrm{\mu m}$ was chosen
to enable octave-spanning DKS frequency comb generation in the resonator's
$\mathrm{TE_{00}}$ mode family for a $1.3\,\mathrm{\mu m}$ pump
laser. The simulation predicts dispersive wave formation around $1\,\mathrm{\mu m}$
and $2\,\mathrm{\mu m}$ for a $1.425\,\mathrm{\mu m}$ wide, fully
$\mathcal{\mathrm{SiO_{2}}}$ cladded waveguide with 82\textdegree{}
sidewall angle. As shown in Figure \ref{figWaferDispersion}(a), a
waveguide height deviation of only $10\,\mathrm{nm}$ causes a shift
of the low frequency dispersive wave position by $5\,\mathrm{THz}$
(i.e. ca. $\Delta\lambda=100\,\mathrm{nm}$ in wavelength). A change
in the waveguide width does not strongly influence the position of
the low frequency dispersive wave but changes the position of the
high frequency dispersive wave. 

The dispersion of different samples is evaluated based on the spectral
envelopes of the generated frequency combs. The low values of microresonator
GVD (i.e. $D_{2}/2\pi\sim\mathcal{O}(20\,\mathrm{MHz)}$ ) and the
bandwidth limitations of our precision diode laser spectroscopy setup
\citep{Liu2016}, did not allow to measure higher order dispersion
terms. Figure \ref{figCoherence}(a) shows the setup used for DKS
frequency comb generation. A $1.3\,\mathrm{\mu m}$ external cavity
diode laser (ECDL) is amplified using a tapered semiconductor amplifier
(SOA) to a maximum power of $0.75\,\mathrm{W}$. Laser light is coupled
into the photonic chip using lensed fibers and inverse tapered waveguides
with about $\sim3\,\mathrm{dB}$ loss per facet. During the laser
tuning into the $\mathrm{TE_{00}}$ mode family resonance, the transmission
through the device as well as the generated comb light are recorded
using photodiodes ($f_{\mathrm{3}\mathrm{dB}}=1\,\mathrm{GHz}$).
Up to three optical spectrum analyzers (OSA) are used to capture
the generated broadband comb spectra. Figure \ref{figWaferDispersion}(b)
shows octave spanning unstable modulation instability (uMI) comb states
excited in three different microresonator chips. The sample position
on the 4'' wafer is provided as inset. As can be seen, the position
variation of the low frequency dispersive wave for the three samples
is $5.5\,\mathrm{THz}$ around to the designed target value of $150\mathrm{\,THz}$.
In comparison with the simulations in Figure \ref{figWaferDispersion}(a)
this indicates a variation of less than $\pm10\,\mathrm{nm}$ in waveguide
height around the target value of $0.74\mathrm{\,\mu m}$. Moreover,
the width variation allows the position of the high frequency dispersive
wave to be tuned. A waveguide width reduction of $50\,\mathrm{nm}$
in the design moves the position of the high frequency dispersive
wave by $\sim25\,\mathrm{THz}$ in good agreement with simulations.
These measurements reveal \emph{actual} lithographic control over
the dispersive wave position, rather than post fabrication selection.
The photonic Damascene process with optimized planarization step thus
offers sufficient precision in the control over waveguide dimensions
to allow for wafer-scale fabrication of microresonators with engineered
dispersion properties over a full octave of bandwidth. 

\section{Coherence and soliton identification}

\begin{figure*}
\includegraphics{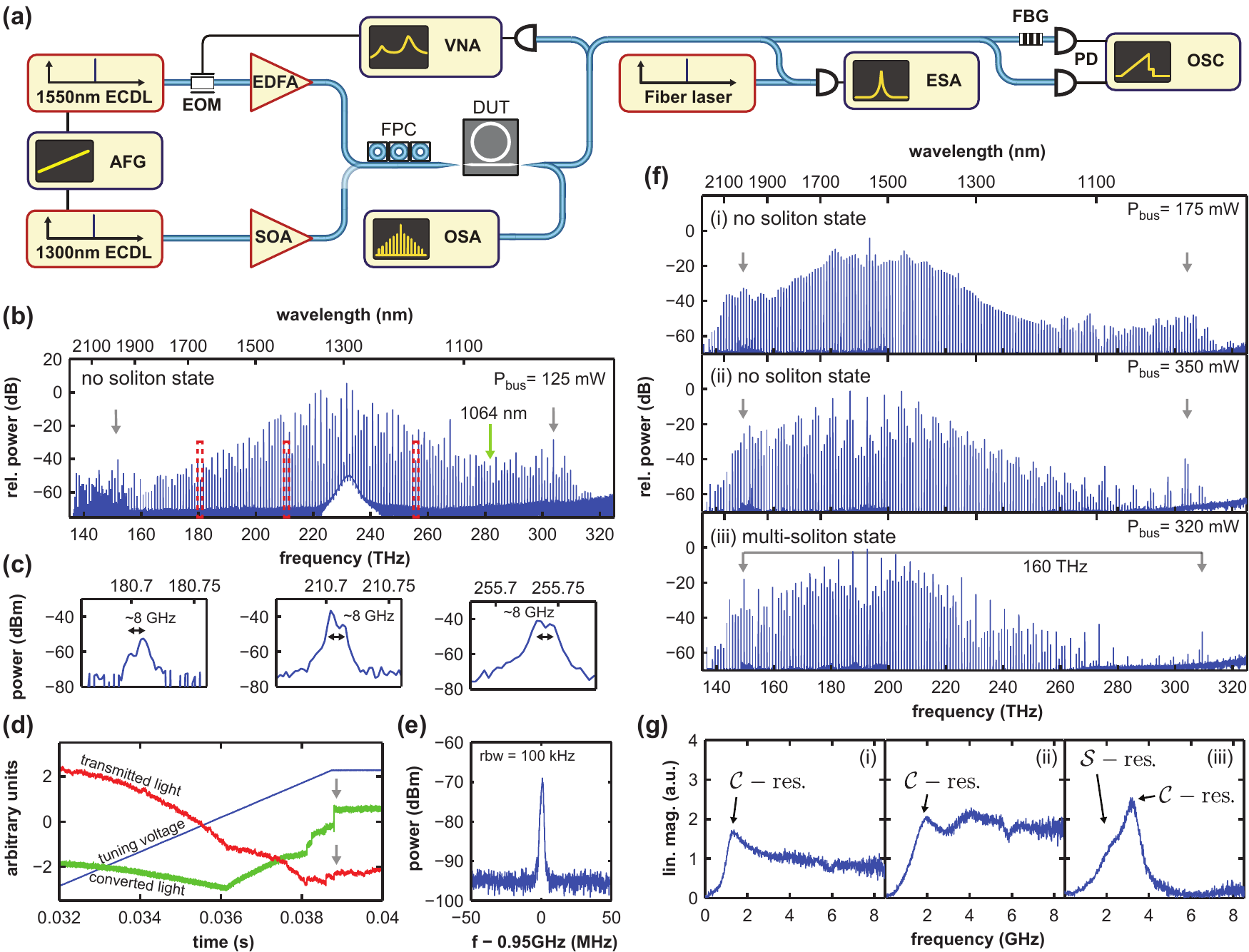}

\caption{Distinction of octave-spanning multi-soliton states and non-solitonic
states. (a) Schematic of the setup used to generate octave-spanning
Kerr combs using $1.3\,\mathrm{\mu m}$ or $1.55\,\mathrm{\mu m}$
pump lasers. In both cases an arbitrary waveform generator (AFG) provides
the voltage ramp to tune the external cavity diode seed laser (ECDL)
into a resonance of the device under test (DUT). Either an erbium-doped
fiber amplifier (EDFA) or a semiconductor tapered amplifier (SOA)
is used to amplify the seed laser before coupling onto the photonic
chip using lensed fibers. An oscilloscope (OSC) records the transmitted
and converted comb light power during the scan. Several optical spectrum
analyzers (OSA) are used to capture the full spectra of the excited
comb state. A $1064\mathrm{\,nm}$ fiber laser allows the recording
of heterodyne beat-notes using a fast photodiode and an electrical
spectrum analyzer (ESA). The comb state response is measured by a
vector network analyzer (VNA) that drives an electro-optic phase modulator
(EOM) and receives part of the transmitted light on a photodiode with
$25\,\mathrm{GHz}$ bandwidth. (b) Highly structured octave-spanning
comb state excited by tuning the $1.3\mathrm{\,\mu m}$ pump laser
into the step-like feature (which, importantly, does not originate
from soliton formation) highlighted in (d) with gray arrows. The red
boxes mark comb teeth which reveal a splitting of $\sim8\,\mathrm{GHz}$
upon close examination in (c), demonstrating subcomb formation. (d)
Oscilloscope trace of the voltage ramp (blue), the generated comb
light (green) and the transmitted light signal (red) (e) Narrow heterodyne
beat note of a $1064\,\mathrm{nm}$ fiber laser with the comb tooth
marked with a green arrow in (b) of a non-solitonic comb state. (f)
Three comb states excited in three different microresonators with
the same waveguide dimensions. (g) Response measurements associated
with the comb states in (f). The position of the cavity ($\mathcal{C}$-res.)
and the soliton ($\mathcal{S}$-res.) resonance is indicated.\label{figCoherence}}
\end{figure*}

For applications of the octave-spanning Kerr comb state coherence
is a key requirement. Common methods to classify the coherence of
Kerr comb states are the measurements of the low frequency amplitude
noise of the generated comb light, the repetition rate beat-note,
as well as heterodyne beat-notes with reference lasers \citep{Herr2012}.
Among the different low noise comb states \citep{Saha:13,Xue2015,Herr2012},
the DKS state is particularly attractive as it provides high coherence
across its entire bandwidth. Soliton Kerr comb states have in particular
been identified via the characteristic transmission steps in the
generated comb light, via the red-detuned nature of the pump laser,
as well as the characteristic soliton spectral envelope. Although
the spectra contain no direct information on coherence, for microresonators
with dominant (positive) quadratic dispersion $D_{2}$ the single
soliton state exhibits a characteristic spectral $\mathrm{sech}^{2}$
envelope. For broadband Kerr combs the spectral envelope can exhibit
dispersive waves, due to higher order microresonator dispersion.
Although dispersive wave features are also present in the uMI state
spectra, in the DKS state their width reduces with a simultaneous
increase in the individual comb tooth power and a shift of the dispersive
wave maximum \citep{Brasch2016}. Other spectral signatures of the
DKS regime include the soliton Raman self-frequency shift, that leads
to a redshift of the soliton's $\mathrm{sech}^{2}$ envelope with
respect to the pump \citep{Karpov2016}. The Raman self-frequency
shift is absent in the uMI comb state, but can be masked in the DKS
state by the soliton recoil due to dispersive wave formation. For
octave spanning soliton Kerr combs with THz mode spacing and dispersive
wave emission due to microresonator GVD and avoided modal crossings
\citep{Yang2016a}, DKS state identification purely based on the
spectral envelope is unreliable.

In contrast, a recently introduced response measurement technique
\citep{Guo2016} allows to unambiguously identify DKS comb states,
and is applied here for the first time to octave spanning soliton
states. Figure \ref{figCoherence}(a) shows the setup used to investigate
the coherence of comb states generated with a $1.3\,\mathrm{\mu m}$
and $1.55\,\mathrm{\mu m}$ ECDL. The $1.55\,\mathrm{\mu m}$ ECDL
is amplified using an erbium-doped fiber amplifier (EDFA) and in order
to perform the response measurement, a phase modulator with $10\,\mathrm{GHz}$
bandwidth is added before the amplifier. A vector network analyzer
(VNA), probing the Kerr comb state's response, drives the EOM and
receives the signal via a photodiode with $25\,\mathrm{GHz}$ bandwidth.
We note that this measurement was only possible using the $1.55\,\mathrm{\mu m}$
pump laser due to the lack of an appropriate phase modulator for the
$1.3\mathrm{\,\mu m}$ pump path. A heterodyne beat note of the comb
with a fiber laser at $1064\,\mathrm{nm}$ is recorded using an electrical
spectrum analyzer (ESA). All presented Kerr frequency comb states
are excited using a simple laser tuning method, applying a linear
voltage ramp provided by an arbitrary function generator (AFG) to
controllably tune into a comb state \citep{Herr2013a,Guo2016}.

DKS formation is accompanied by characteristic step features that
occur simultaneously in the generated comb and transmitted light trace
\citep{Herr2013a}. However, in the experiments reported here in Figure
\ref{figCoherence}(b)-(d), we found that abrupt changes in the transmitted
power do\emph{ not} necessarily originate from soliton formation.
Instead, abrupt changes between different non-solitonic comb states,
as well as resonance splitting due to waveguide surface roughness
or near-by resonances of other mode families can cause similar features.
Figure \ref{figCoherence}(b) shows the optical spectrum of a comb
state generated by tuning the $1.3\,\mathrm{\mu m}$ pump laser with
$\mathrm{P_{bus}}=125\,\mathrm{mW}$ in the bus waveguide into the
step feature visible in Figure \ref{figCoherence}(d). The spectrum
spans an octave and is highly structured, featuring individual sharp
lines at the dispersive wave positions. Moreover, the beat-note of
an individual comb tooth with a fiber laser at $1064\,\mathrm{nm}$
(see Figure \ref{figCoherence}(e)) shows low noise characteristics
i.e. $>20\mathrm{\,dB}$ signal-to-noise ratio in a $100\,\mathrm{kHz}$
resolution bandwidth (RBW). Based on these measurements, the generated
comb state would agree with properties associated with a multi-soliton
state. However, close examination of the generated spectral lines,
shown in Figure \ref{figCoherence}(c), reveals that certain comb
lines have a splitting of $\sim8\,\mathrm{GHz}$, \textit{incompatible}
with a DKS comb state and associated with merging of individual \textit{subcombs}
\citep{Herr2012}. Importantly, commonly employed \citep{Herr2012,Li2016}
low frequency amplitude noise measurements, and even local heterodyne
beat-note measurements, are insufficient discriminators, unless they
are performed over sufficiently wide bandwidth to detect the subcomb
spacing (here $\gg1\,\mathrm{GHz}$). Indeed, for the measurements
above, no noise $<1\mathrm{\,GHz}$ is observed, which would lead
to an erroneous soliton state identification. 

A more reliable method is using the response of the comb state to
a phase modulation of the pump laser. It was shown that the phase
modulation response function of DKS states consists of two characteristic
resonances originating from the circulating soliton pulses ($\mathcal{S}$-resonance)
and the cavity response ($\mathcal{C}$-resonance) \citep{Guo2016,Lucas2016a}.
Furthermore, the measurement allows to infer the effective pump laser
detuning within the DKS state, an important quantity determining the
soliton properties. Figure \ref{figCoherence}(f) shows three comb
states recorded from three microresonators on the same photonic chip
having the same waveguide dimensions but varying bus-resonator distances.
Again, the microresonator dispersion was designed for dispersive wave
emission around $1\,\mathrm{\mu m}$ and $2\,\mathrm{\mu m}$, here
upon pumping with a $1.55\,\mathrm{\mu m}$ laser. We excite different
comb states and analyze their corresponding response functions shown
in Figure \ref{figCoherence}(g). State (i) is a uMI state and the
response function exhibits a single peak and a non-zero background
noise at high frequencies. State (ii) has a much more structured spectral
envelope and a seemingly sharp dispersive wave feature compared to
state (i). However, the response measurement reveals a fundamentally
different response than expected for a DKS state. Also, state (iii)
has a strongly structured spectral envelope with similarities to the
envelope of state (ii) but its response function shows two overlapping
resonances. The lower frequency $\mathcal{S}$-resonance originates
from the circulating soliton pulses and has a reduced amplitude compared
to the cavity $\mathcal{C}$-resonance at higher frequency, whose
frequency indicates the detuning. Moreover, the separation of both
peaks diminishes upon laser blue detuning and increases for red detuning,
respectively. We thus identify state (iii) as a multi-soliton state.
The multi-soliton state leads to an increased conversion efficiency
compared to the single soliton case, resulting in negligible residual
pump power in the present case. Moreover, we observe a notable shift
of the high frequency dispersive wave peak position (indicated by
gray arrows).

\section{Octave spanning spectra via single soliton generation}

\begin{figure}
\includegraphics{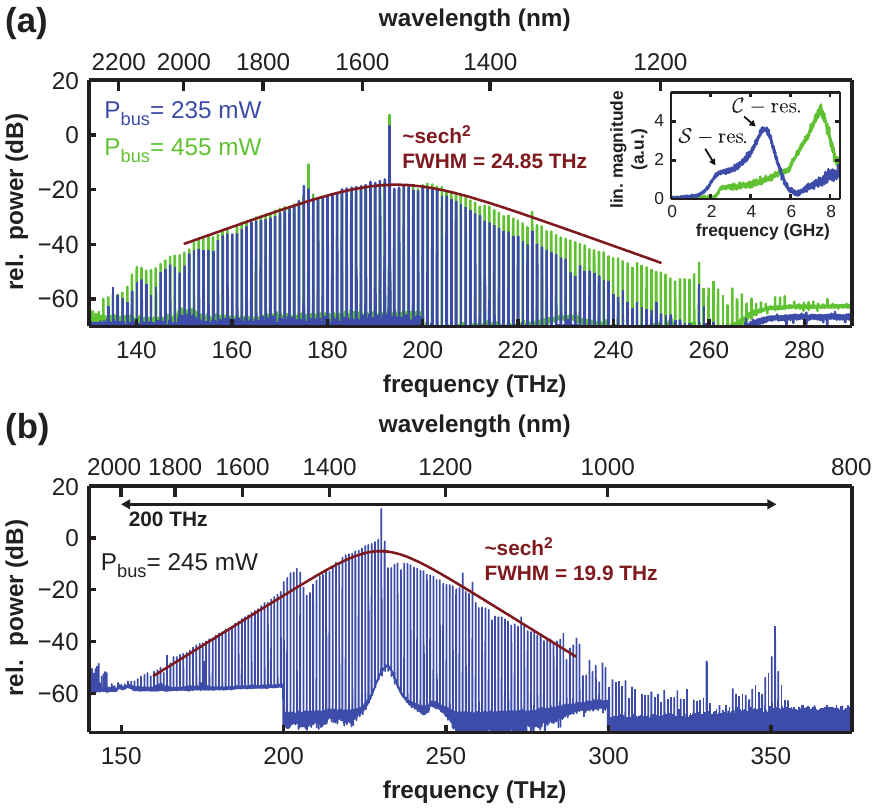}

\caption{Octave-spanning single soliton generation at $1.55\mathrm{\,\mu m}$
and $1.3\mathrm{\,\mu m}$ pump wavelength. (a) The same single DKS
state shown for two different pump laser powers and cavity detunings.
Inset: Response measurement of the single DKS states shown in (a)
revealing the respective pump laser cavity detuning. The $\mathcal{S}$-
and $\mathcal{C}$-resonances are separated by $\sim2.5\,\mathrm{GHz}$
and $\sim5\,\mathrm{GHz}$, respectively. (b) Single DKS spanning
$200\,\mathrm{THz}$ in optical bandwidth with a dispersive wave at
$850\mathrm{\,nm}$ excited using a $1.3\mathrm{\,\mu m}$ pump laser.
Fits of both states with a spectral $\mathrm{sech}^{2}$ envelope
are shown in red. Pump powers in the bus waveguide and fitted spectral
bandwidths are noted. \label{figBestPerformance}}

\end{figure}

Once excited, a multi-soliton state can often be converted into a
single soliton state by soliton switching upon ``backward tuning''
of the pump laser \citep{Guo2016}. Figure \ref{figBestPerformance}
shows two examples of octave-spanning single soliton generation obtained
via this technique using $1.3\mathrm{\,\mu m}$ or $1.55\,\mathrm{\mu m}$
pump lasers. We note that not all multi-soliton states excited were
observed to switch to a lower soliton number upon backward tuning,
but some changed into non-solitonic states. Moreover, it is observed
that both single soliton combs are excited in direct vicinity of an
avoided modal crossing causing strong local deviations from the characteristic
$\mathrm{sech}^{2}$ shaped spectral envelope \citep{Herr2013}.

The soliton spectrum shown in Figure \ref{figBestPerformance}(a)
features no dispersive waves indicating a dominant quadratic factor
of the anomalous GVD. The response measurement shown in the inset
shows a double resonance signature and reveals a cavity detuning of
$\sim4.5\,\mathrm{GHz}$. The pump power is increased to $455\,\mathrm{mW}$
in the bus waveguide while maintaining the single DKS state, which
allows for a larger soliton existence range and octave bandwidth at
a cavity detuning of $\sim7\,\mathrm{GHz}$. This cavity detuning
is significantly higher than previously published values for $\mathrm{Si_{3}N_{4}}$
microresonators indicating a strong nonlinear phase shift due to the
high peak intensity of the soliton pulse \citep{Guo2016}. The spectral
envelope is fitted using a $\mathrm{sech}^{2}$ shape and a $3\,\mathrm{dB}$
bandwidth of $24.85\,\mathrm{THz}$ is extracted, corresponding to
a $12.7\,\mathrm{fs}$ pulse. In Figure \ref{figBestPerformance}(b)
we present the \textit{broadest }single DKS comb state published to
date, to the best of our knowledge. The DKS comb spans a total bandwidth
of $\sim200\,\mathrm{THz}$ using a pump laser at $1.3\,\mathrm{\mu m}$
providing only $245\,\mathrm{mW}$ power in the bus waveguide. The
fitted $3\mathrm{\,dB}$ bandwidth of $19.9\mathrm{\,THz}$ is similar
to the value of the single DKS state shown in Figure \ref{figBestPerformance}(a)
even though the required pump power is significantly lower. This demonstrates
the strong influence of dispersion on the power requirements for broadband
comb generation and underlines the necessity for precise dispersion
control.

\section{Conclusion}

In conclusion, we have demonstrated DKS generation in $\mathrm{Si_{3}N_{4}}$
microresonators with $1\,\mathrm{THz}$ FSR and spectral bandwidths
exceeding one octave, both at $1.55\,\mathrm{\mu m}$ and $1.3\,\mathrm{\mu m}$
pump wavelength. The photonic Damascene process with optimized planarization
step allows for high yield, wafer-scale fabrication of microresonators
with precisely controlled dispersive wave positions. Our work moreover
revealed that for comb states generated in $\mathrm{Si_{3}N_{4}}$
microresonators with $1\,\mathrm{THz}$ FSR, conventional criteria
of coherence can fail. We carefully investigated the coherence of
generated comb states and unambiguously identified octave-spanning
multi-DKS states based on their unique phase modulation response signature.
Finally, we demonstrated single soliton generation with record bandwidth
of $200\,\mathrm{THz}$ generated with only $245\,\mathrm{mW}$ of
pump power. 

Our findings demonstrate the technological readiness of the photonic
Damascene process and integrated $\mathrm{Si_{3}N_{4}}$ frequency
comb generators for a wide range of applications. The ability to generate
ultra-short pulses also with a $1.3\,\mathrm{\mu m}$ pump laser,
as demonstrated here, opens up applications in biological and medical
imaging as this wavelength represents a compromise between low tissue
scattering and absorption due to water. Moreover, such DKS frequency
combs with designed spectral envelopes may in the future enable chip
integration and miniaturization of self-referenced frequency combs
\citep{Bowers2016} and dual comb CARS \citep{Ideguchi2013}.

\subsection*{Acknowledgements}

$\mathrm{Si_{3}N_{4}}$ microresonator samples were fabricated in
the Center of MicroNanoTechnology (CMi) at EPFL.

This publication was supported by Contract HR0011-15-C-0055 from the Defense Advanced Research Projects Agency
(DARPA), Defense Sciences Office (DSO), funding from Air Force Office of Scientific Research, Air Force Material Command, USAF under Award No. FA9550-15-1-0099, the European
Union\textquoteright s Horizon 2020 research and innovation programme
under MIRCOMB: Marie Sklodowska-Curie IF grant agreement No. 709249
and the Swiss National Science Foundation under grant agreement No.161573
and 163864. M.K. acknowledges the support from the Marie Curie Initial
Training Network FACT.

This research was carried out concurrently
with the work from K. S. at NIST in the framework of the project DODOS
that is available online under \citep{Li2016} and also reports octave
spanning Kerr soliton combs with dual dispersive waves.

\bibliographystyle{apsrev4-1}
\bibliography{thzCombBibliography}

\end{document}